\documentclass[final,5p,times,twocolumn]{elsarticle}

\usepackage{amsmath}
\usepackage{amssymb}
\usepackage{bm}
\usepackage{mathrsfs}

\usepackage{graphicx}
\usepackage{lscape}
\usepackage{epstopdf}
\usepackage{subfigure}
\usepackage{bmpsize}

\usepackage{natbib}
\usepackage{pxfonts}

\usepackage{xr}
\usepackage{color}

\usepackage{multirow}
\usepackage{booktabs}  
\usepackage{threeparttable}

\usepackage{fancyhdr}
\usepackage[hang,flushmargin]{footmisc}

\usepackage[colorlinks, linkcolor=blue, anchorcolor=blue, citecolor=blue]{hyperref}

\journal{}

\begin{document}
	\begin{frontmatter}
		
		\title{\textbf{A modified formula for non-Arrhenius diffusion of helium in metals}}
		
		\author[label_IFCEN,label_PhysDep]{Haohua Wen}
		\author[label_IFCEN,label_PhysDep]{Kan Lai}
		\author[label_IFCEN,label_PhysDep]{Jianyi Liu}
		\author[label_PhysDep]{Yifeng Wu}
		
		
		\author[label_PhysDep,label_KeyLab]{Yue Zheng\corref{CA1}}
		\cortext[CA1]{Corresponding author: zhengy35@mail.sysu.edu.cn}
		
		\address[label_IFCEN]{
			Sino-French Institute of Nuclear Engineering and Technology, Sun Yat-sen University, Zhuhai 519082, China
		}
	
		\address[label_PhysDep]{
			Micro\&Nano Physics and Mechanics
			Research Laboratory, School of Physics, Sun Yat-sen University, Guangzhou 510275, China
		}
		
		\address[label_KeyLab]{
			State Key Laboratory of Optoelectronic Materials and Technologies, School of Physics, Sun Yat-sen University, Guangzhou 510275, China
		}
	
\begin{abstract}
	Helium diffusion in metals is the basic requirement of nucleation and growth of bubble, which gives rise to adverse degradation effects on mechanical properties of structural materials in reactors under irradiation. Lattice based Kinetic Monte Carlo approach is widely adopted to study the evolution of helium-vacancy clustering. However, the implementation of Arrhenius law to prediction the event rate of single interstitial helium solute diffusion in metal is not always appropriate due to low-energy barrier. Based on a stochastic model, a modified formula is derived from the Brownian motion upon a cosine-type potential. Using the parameters obtained from molecular dynamics simulation for the diffusivity of single helium solute in BCC W, the prediction of our model is consistent with the results from dynamical simulation and previous model. This work would help to develop a more accurate KMC scheme for the growth of helium-vacancy clusters, as well as other low-energy reactions in materials science.   
\end{abstract}


\end{frontmatter}
		
\section{\label{Sec.1}{Introduction}}

	Helium (He) is one of the most common productions in structural materials of fission and fusion reactors under irradiation. Due to the insolubility of He atom in metals, it would be easily trapped into sinks, such as vacancies and grain-boundary, and would form the helium-bubble in a long-term evolution process. The formation and accumulation of helium plays important role in the high-temperature kinetics of microstructure evolution of nuclear materials under irradiation, leading to the adverse ageing effects, e.g. high-temperature helium embrittlement  \cite{duffy2011modelling,zinkle2013materials}. Therefore, the mechanistic understanding of bubble evolution is considered as one of the key issues in nuclear materials science and engineering \cite{TRINKAUS2003229}. In this paper, we concentrate on the non-Arrhenius diffusion behavior of helium in metals. 
	
\begin{figure}
		\graphicspath{}
		\makeatletter
		\def\@captype{figure}
		\makeatother
		\centering
		\includegraphics[width=0.45\textwidth]{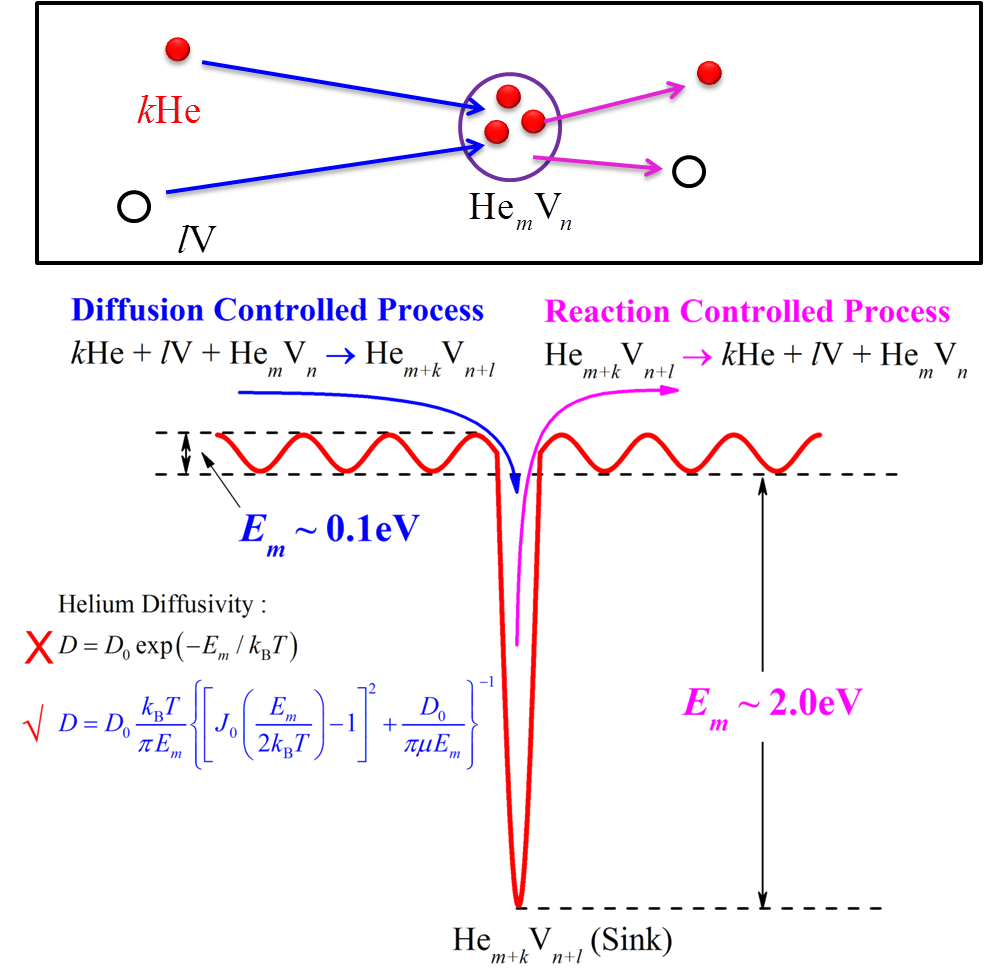}
		\caption{A schematics of the evolution of helium-vacancy clustering treated as a chemical-reaction-like process (See in Eq.~(\ref{Eq.:chemical_reaction})) in KMC approach, where the forward reaction is diffusion-controlled by means of helium (red solid-dots in upper figure) migrates towards the sink (violet open-circle with red solid-dots inside), i.e., $\textrm{He}_m\textrm{V}_n$, indicated by the energy profile (red solid-line in the lower figure).}
		\label{Fig.1}
	\end{figure}

	As long-term phenomena, the growth of helium bubbles in metals are made up by numerous atomic activation processes, including the migration of defects, the combination of helium and vacancies, and the dissociation of helium-vacancy clusters \cite{samaras_multiscale_2009}. In theoretical research, a well-developed \emph{lattice based kinetic Monte Carlo} (short for ``KMC" in the rest of paper) approach \cite{bortz1975a,fichthorn1991theoretical} is widely adopted, by treating the evolution of helium-vacancy clusters as a chemical-reaction-like process, e.g., 
		\begin{equation} \label{Eq.:chemical_reaction}
			 k\textrm{He} + l\textrm{V} + \textrm{He}_m\textrm{V}_n \rightleftharpoons 	\textrm{He}_{m+k}\textrm{V}_{n+l}
		\end{equation}
	as shown in Fig.~\ref{Fig.1}. Further, each individual atomic process is treated as a classical reaction event, whose occurring frequency $\nu$, i.e., event rate, is associated with the reaction-path and the corresponding energy barrier $E_m$, usually described by Arrhenius equation in KMC \cite{voter_book_2007}, as
		\begin{equation} \label{Eq.:event_rate}
			\nu = \nu_0 \exp(-E_m/k_\textrm{B}T)
		\end{equation}
	with $\nu_0$ the attempt frequency, $k_\textrm{B}$ the Boltzmann constant, and $T$ the absolute temperature. 
	
	Note that, because of the repulsive interaction between helium and metal atoms and a small migratory energy barrier, the forward reaction process in Eq.~(\ref{Eq.:chemical_reaction}) seems to be diffusion-controlled at finite temperatures (See Fig. 1 of Ref.~\cite{yuelin2013trapping} or the schematics in Fig.~\ref{Fig.1}), that the growth rate of a helium-vacancy cluster is mainly determined by the mass-flux of helium towards the trap \cite{trocellier_review_2014}, equivalent to the diffusion behavior. In the present KMC scheme, the diffusivity $D$ is usually described by an Arrhenius equation, as
		\begin{equation}\label{Eq.:D_AR}
			D = g\lambda^2\nu = D_0\exp(-E_m/k_\textrm{B}T)
		\end{equation}
	where $D_0 = g\nu_0\lambda^2$ is a pre-factor, with $g$ a geometrical factor and $\lambda$ the distance between two adjunct stable sites in real-space. 

	For the thermal assistant diffusion event, the prerequisite of using Arrhenius equation in Eq.~(\ref{Eq.:D_AR}) to estimate the diffusivity is that the diffusion should be treated as a thermodynamic reversible process, where the energy barrier between two adjunct stable sites along the reaction path should be far larger than the thermal energy, i.e., $E_m \gg k_\textrm{B}T$. Equivalently, the mean time for the defect escaping out of the trap should be far larger than the characteristic relaxation time for the defect-system from saddle point configuration towards equilibrium state. For the helium-vacancy clusters or mono-vacancy, the migration energy is in the order of $\sim$1eV, which satisfies this prerequisite in a wide range of temperature, e.g., $T < 10^4$K. In this case, the diffusion behavior could be well-described by Arrhenius equation \cite{TRINKAUS2003229}.  
	
	However, calculations based on density functional theory (DFT) \cite{PhysRevLett.97.196402} indicated that $E_m\sim$ 0.1eV of single interstitial helium in metals, therefore, the migratory energy barrier is comparable with (even larger than), the thermal energy at $T > 1000$K, breaking down the condition of $E_m \gg k_\textrm{B}T$. In this scenario, it will reveal anomalous diffusion behavior beyond the Arrhenius description \cite{dudarev2008non}, as demonstrated by the recent molecular dynamics (MD) simulations \cite{wen_many-body_2017, perez2014diffusion, SHU201384} of helium diffusion in BCC W ($E_m \sim 0.15$eV). More precisely, the real-space trajectory of interstitial helium at 2600K (See Fig. 2(b) in Ref.~\cite{SHU201384}) indicated that it seems not a hopping random walker with a long resident time between its stable sites, but a thermal assisted Brownian motion in a stochastic media. Therefore, the high-temperature diffusivity of single interstitial helium in BCC W does not obey Arrhenius equation but the Einstein-Smoluchowski equation, i.e., 
		\begin{equation}\label{Eq.:ED_equation}
			D = \mu k_\textrm{B}T = \frac{k_\textrm{B}T}{m^*\gamma} \propto T
		\end{equation}
	where $\mu = 1/m^*\gamma$ is the classical mobility, with $m^*$ the effective mass of the diffusing object, and $\gamma$ the friction coefficient provided by the host media for the diffusion. 
	
	In fact, the diffusion behavior of defects having a small transition barrier, e.g., $E_m \leqslant 0.1 $eV, is categorized to be the low-energy reactions \cite{henkelman_atomistic_2017}, which is found to be a common phenomenon in other disciplines, such as adatom migration upon crystal surface \cite{ALANISSILA1992227, lechner2013atomic,krylov2014physics}, self-interstitial diffusion \cite{dudarev2008non} and dislocation motion \cite{PhysRevB.84.134109} in metals. Following Kramers' theory \cite{KRAMERS1940284}, theoretical studies \cite{sancho2004diffusion, pavliotis2008diffusive} discussed the physical picture of these low-energy reactions: (1) the low-temperature behavior is governed by Arrhenius equation at $E_m \gg k_\textrm{B}T$, which the quasi-equilibrium properties, i.e., $E_m$ and $\nu_0$ shown in Eq.~(\ref{Eq.:event_rate}), are used to described the reversible nature; (2) the high-temperature behavior is governed by Einstein-Smoluchowski equation at $E_m \ll k_\textrm{B}T$, which the non-equilibrium properties, like $\mu$ shown in Eq.~(\ref{Eq.:ED_equation}) to determine the heat dissipation, are used to describe the irreversible nature \cite{dudarev2002thermal,DUDAREV2002881,Swinburne2014}. On the basis of this physical picture, attention have been paid on the low-energy reaction behavior at $E_m \sim k_\textrm{B}T$, the intermediate stage between the low- and high-temperature limits, by integrating both the reversible and irreversible characteristics into a unique framework \cite{sancho2004diffusion, pavliotis2008diffusive}, which is believed to be the key issue for the theoretical description of the nature of the low-energy reaction.
	
	For the issue of the nucleation and growth of helium-vacancy clusters, most of the multi-scale modeling studies based on KMC focus on the accuracy of input parameters calculated from DFT or MD simulations \cite{samaras_multiscale_2009}, rather than the more important issue, i.e., to introduce the low-energy reaction mechanism into the KMC scheme\cite{henkelman_atomistic_2017}. One of the recent endeavors is the work of Ref.~\cite{wen_many-body_2017} (short for ``saw-tooth model" in the following), who used Zwanzig \cite{Zwanzig1960} and Mori \cite{Mori1965} projection operator approach to construct a stochastic model of Brownian motion upon a simplified saw-tooth potential and propose a unique equation to describe the `\emph{anomalous}' diffusion behavior for single helium in BCC W in the whole temperature region (See Fig. 4 in Ref.~\cite{wen_many-body_2017}). However, the saw-tooth model is derived on the basis of Brownian motion, where the diffusion behavior at the intermediate temperature region is described by the simple interpolation of the behaviors at low- (i.e., Eq.~(\ref{Eq.:D_AR})) and high-temperature limits (i.e., Eq.~(\ref{Eq.:ED_equation})). Further, the thermal fluctuation information at the basin of potential well for helium diffusion is missed due to the usage of simple saw-tooth potential. 
	
	To overcome the limitations in saw-tooth model \cite{wen_many-body_2017}, we construct an alternative stochastic model of hopping random walk upon a cosine potential (short for ``cosine model" in the following), then derive a modified formula of single helium diffusion in metals, which is suitable for the KMC scheme in the issue of growth of helium-vacancy clusters. As presented in Sec.~\ref{Sec.2}, cosine model will start with an assumption that the stochastic motion of single helium in metals could still be treated act as a step-by-step random walker hopping among its stable sites, i.e., the basins of a cosine potential provided by atoms of the host metal, and the motion of helium around its potential minimum is regarded as a damped harmonic motion, upon which an \emph{effective quality factor} is introduced to describe the `anomalous' diffusion behavior at the intermediate temperatures. The comparison of prediction of cosine model and the results obtained from molecular dynamics simulations will be then placed in Sec.~\ref{Sec.3}, as well as the discussion. We hope this modified formula could help to develop a more accurate KMC scheme for the long-term evolution of helium-vacancy clusters and bubbles, as well as other low-energy reactions.
	
\section{\label{Sec.2} A modified analytical formula}

	\subsection{Stochastic dynamics model}
	In BCC metals, a single helium atom occupying the tetrahedral interstitial sites generally exerts a biased periodic crystal potential and results in the local distortion and resonance vibrational modes, as well as the scattering center for phonon modes of the host matrix. With the thermal fluctuations and interaction provided by host atoms, helium atom travels upon the crystal potential until it is trapped by sinks, as schematic as in Fig. 1. Considering the case that helium atom is far away from the sinks in BCC W, the migration energy $E_m$ of interstitial helium in BCC W is found as small as $\sim$0.145eV \cite{wen_many-body_2017}, whose diffusion could be regarded as a low-energy reaction. According to saw-tooth model \cite{wen_many-body_2017}, the trajectory of interstitial helium solute inside the metals $X=X(t)$ is governed by a stochastic Langevin equation as
		\begin{equation} \label{Eq.:Langevin Equation}
			m^* \ddot{X} = F - m^*\gamma\dot{X} + f(t)
		\end{equation}
	where $X$ is the instantaneous position of helium atom, with $m^*$ the effective mass; $F = -\partial_X E(X)$ is the restoring force, with $E(X)$ the periodic crystal potential provided by host matrix, i.e., \emph{migratory barrier}; $-m^*\gamma\dot{X}$ and $f(t)$ represent respectively the actions of dissipation and fluctuation provided by host atoms, with $\gamma$ the phonon-drag friction coefficient and $f(t)$ a Gaussian-type random force, where $\left\langle f(t) \right\rangle = 0$, and $\left\langle f(t)f(t')\right\rangle = 2m^*\gamma k_\textrm{B}T\delta(t-t')$. 
	
	The diffusivity $D$ could be derived from the trajectory $X(t)$ governed by Eq.~(\ref{Eq.:Langevin Equation}), as	
		\begin{equation}
			D = \frac{\left\langle \left[ X(t) - X(0) \right]^2 \right\rangle }{2t} = \frac{m^*}{k_\textrm{B}T}\int_{0}^{\infty} {{\left\langle \dot{X}(t) \dot{X}(0) \right\rangle} \textrm{d}t }
		\end{equation}
	\noindent where $\left\langle \cdots \right\rangle $ represents the running-time average. In particular, 
		\begin{equation}
			D = 
			\left\{
				\begin{aligned}
					& D_0 e^{-E_m/k_\textrm{B}T}, & \textrm{if} \ \ E_m \gg k_\textrm{B}T\\
					& \mu k_\textrm{B}T, & \textrm{if} \ \  E_m \ll k_\textrm{B}T \\
				\end{aligned}
			\right.
		\end{equation}
	at low- and high-temperature limits, where $D_0 = g\nu_0\lambda^2 = \nu_0\lambda^2$ with $g=2/3$ for helium diffusion in BCC metals. The problem is that it is very difficult to obtain the analytical expression of $D$ at the intermediate temperature, i.e., $E_m \sim k_\textrm{B}T$, for an arbitrary potential $E(X)$ in Eq.~(\ref{Eq.:Langevin Equation}). Substantial progresses were achieved just in the limiting cases. The representative works are respectively the Lifson-Jackson formula \cite{Lifson-Jackson} in the large friction limit
		\begin{equation}\label{Eq.:Lifson-Jackson}
			D = \mu k_\textrm{B}T \left[  \int_{0}^{\lambda} e^{\frac{E(X)}{k_\textrm{B}T}} \textrm{d}X  \int_{0}^{\lambda} e^{-\frac{E(X)}{k_\textrm{B}T}} \textrm{d}X \right]^{-1}, \quad \textrm{if} \ \  \gamma \gg \nu_0
		\end{equation}		
	\noindent with $\mu = 1/m^*\gamma$ the classical mobility, and Risken's expression \cite{Risken1996} based on a cosine-type potential in the low friction limit
		\begin{equation}\label{Eq.:Risken}
			D = \frac{\pi k_\textrm{B}T}{\gamma E_m} \exp(-E_m/k_\textrm{B}T), \quad \textrm{if} \ \  \gamma \ll \nu_0
		\end{equation}	
	\noindent In underdamped limit, i.e., $\gamma \gg \nu_0$, the defect trajectory shows the long tracks ($\gg$$\lambda$) inside the metals, otherwise the overdamped condition, i.e.,  $\gamma \ll \nu_0$, leads to a typical short steps ($\sim$$\lambda$) of hopping between the potential minima \cite{sancho2004diffusion}. The former case is suitable to describe the high-temperature behavior of helium diffusion \cite{perez2014diffusion}, but the form of the latter case is more applicable in KMC scheme. Therefore, we would like to derive our analytical expression on the basis of Eq.~(\ref{Eq.:Lifson-Jackson}) and introduce an \emph{effective quality factor} $b=g \cdot \left( \gamma/2\pi\nu_0 \right) $, to account for the damping feature for helium diffusion upon a cosine potential in the following.

	\subsection{Derivation of cosine model}	
	Assuming a cosine-type force field of $E(X)$ is exerted by a single interstitial helium in metals as
		\begin{equation}\label{Eq.:cosine_potential}
			E(X) = \frac{E_m}{2}\left[ 1 - \cos(2\pi X/\lambda) \right] 
		\end{equation}
	The attempt frequency $\nu_0$ could be then derived as
		\begin{equation}
			4\pi^2\nu_0^2 = \frac{1}{m^*} \left( \frac{\partial^2 E}{\partial X^2}\right)_{X=0}  \Rightarrow 2\pi \nu_0 = \frac{\pi E_m}{m^* \nu_0\lambda^2} =  \frac{\pi g E_m}{m^* D_0}
		\end{equation}
	Therefore, the \emph{quality factor} $b$ is written as
		\begin{equation}
			b =g \frac{\gamma}{2\pi \nu_0} = \frac{g}{m^* \mu} \frac{m^* D_0}{\pi g E_m}
				= \frac{D_0}{\pi \mu E_m}
		\end{equation}
	
		\begin{figure}
			\graphicspath{}
			\makeatletter
			\def\@captype{figure}
			\makeatother
			\centering
			\includegraphics[width=0.45\textwidth]{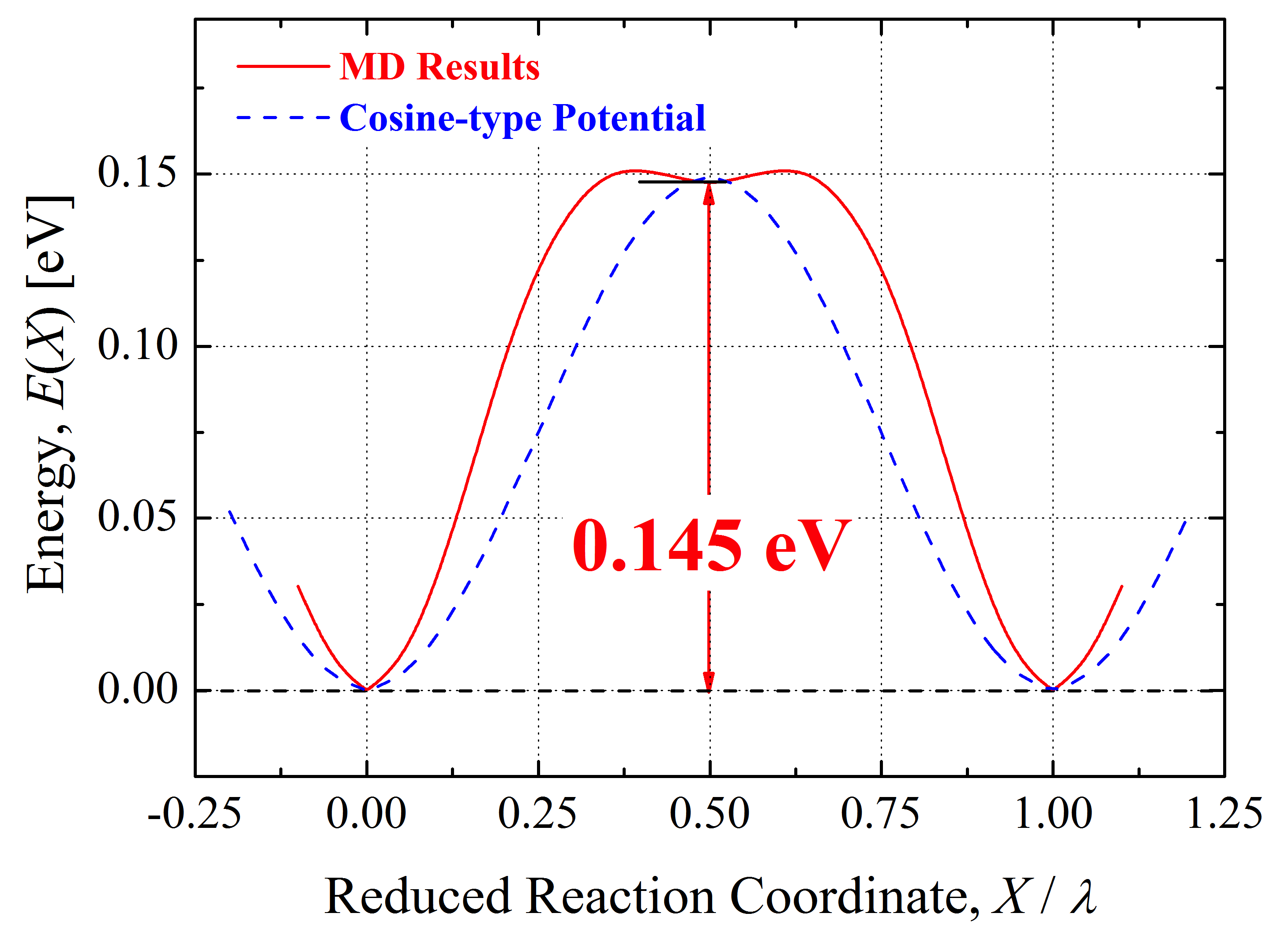}
			\caption{The migratory energy profile of single interstitial helium solute in BCC W hopping between two neighboring tetrahedral sites (red solid-line) obtained using MD-type simulation in Ref.~\cite{wen_many-body_2017}, and the approximate description (blue dashed-line) of a cosine-type potential, i.e., Eq.~(\ref{Eq.:cosine_potential}), with $E_m = 0.145$eV.}
			\label{Fig.2}
		\end{figure}
	
	Meanwhile, the diffusivity could be derived by substituting Eq.~(\ref{Eq.:cosine_potential}) into Eq.~(\ref{Eq.:Lifson-Jackson}), as  \cite{pavliotis2008diffusive}
		\begin{equation} \label{Eq.:D_intermediate}
				D  = \mu k_\textrm{B}T \left[ J_{0}\left( \frac{E_m}{2k_\textrm{B}T}\right)  \right] ^{-2} 
		\end{equation}
	where $J_0(x)$ is the modified Bessel function of the first kind, with $J_0(x)$=1 at $x\ll1$ and $J_0(x)$=$\sqrt{(2\pi x)}e^{x}$ at $x\gg1$. Therefore, 
		\begin{equation} \label{Eq.:D_limit_inter}
			D =
			\left\{
				\begin{aligned}
					& \pi \mu E_me^{-E_m/k_\textrm{B}T }, & \quad \textrm{if} \ \ E_m \gg k_\textrm{B}T \\
					& \mu k_\textrm{B}T \ , & \quad \textrm{if} \ \ E_m \ll k_\textrm{B}T   \\
				\end{aligned}
			\right.
		\end{equation}
	which satisfies the high-temperature limiting behavior but not the one at low-temperature limit. To solve this problem, we modify Eq.~(\ref{Eq.:D_intermediate}) by introducing the \emph{quality factor} $b$, in form of 
			\begin{equation}\label{Eq.:modified_D}
				\begin{aligned}
				D & = \mu k_\textrm{B}T b \left\lbrace \left[ J_0\left( \frac{E_m}{2k_\textrm{B}T}\right)  - 1\right]^2 + b \right\rbrace^{-1} \\
				   & = D_0 \frac{k_\textrm{B}T}{\pi E_m} \left\lbrace \left[ J_0\left( \frac{E_m}{2k_\textrm{B}T}\right)  - 1\right]^2 + \frac{D_0}{\pi\mu E_m} \right\rbrace^{-1} \\ 
				   \end{aligned}
	\end{equation}
	In this regard, we have
		\begin{equation}
			D = 
				\left\{
					\begin{aligned}
						& D_0 e^{-E_m/k_\textrm{B}T} \ , & \quad \textrm{if} \ \ E_m \gg k_\textrm{B}T \\
						& \mu k_\textrm{B}T  \ ,& \quad \textrm{if} \ \ E_m \ll k_\textrm{B}T \\
					\end{aligned}
				\right.
		\end{equation}
	satisfying the low- and high-temperature limiting conditions. The $\gamma^{-1}$ dependence is could be found in Eq.~(\ref{Eq.:modified_D}), which is consistent with the prediction of a more rigorous theory proposed in Ref.~\cite{sancho2004diffusion}. The occurring rate $\nu$ of single helium hopping event conveniently used in KMC is then derived as
		\begin{equation} \label{Eq.:modified_nu}
			\nu = \frac{D}{g\lambda^2} = \nu_0 \frac{k_\textrm{B}T}{\pi E_m} \left\lbrace \left[ J_0\left( \frac{E_m}{2k_\textrm{B}T}\right)  - 1\right]^2 + \frac{D_0}{\pi\mu E_m}  \right\rbrace^{-1}
		\end{equation}
	so that $\nu = \nu_0$ at $E_m \gg k_\textrm{B}T$ and $\nu = \mu k_\textrm{B}T/\lambda^2$ at $E_m \ll k_\textrm{B}T$, which satisfies the Arrhenius and Einstein-Smoluchowski equations, respectively. 
	
	Eq.~(\ref{Eq.:modified_D}) is the derived analytical expression of the diffusivity for a single helium solute in metals, which bridges the characteristics of quasi-equilibrium (i.e., the pre-factor $D_0$ and migration energy barrier $E_m$) and non-equilibrium (i.e., classical mobility $\mu = 1/m^*\gamma$) in low-energy reaction with an \emph{effective quality factor} $b$. 

\section{\label{Sec.3}Numerical example of helium in BCC W}

	Fig.~\ref{Fig.2} shows the migratory energy profile of single interstitial helium solute in BCC W hopping between two neighboring tetrahedral sites, which is calculated using modified conjugated gradient (MCG) \cite{Woo2003} method in Ref.~\cite{wen_many-body_2017}. Accordingly, a cosine-type potential, i.e., Eq.~(\ref{Eq.:cosine_potential}), is used to describe the detail information with $E_m = 0.145$eV. It can be seen the cosine potential is in general agreement with the migratory profile, including the curvature at the basin of potential well and the intermediate part. In addition, the saddle-point state, i.e., the top of migratory profile, is a meta-stable state but with a tiny barrier, $< 0.01$eV, which could not be reproduced using Eq.~(\ref{Eq.:cosine_potential}). 

		\begin{figure}
			\graphicspath{}
			\makeatletter
			\def\@captype{figure}
			\makeatother
			\centering
			\includegraphics[width=0.45\textwidth]{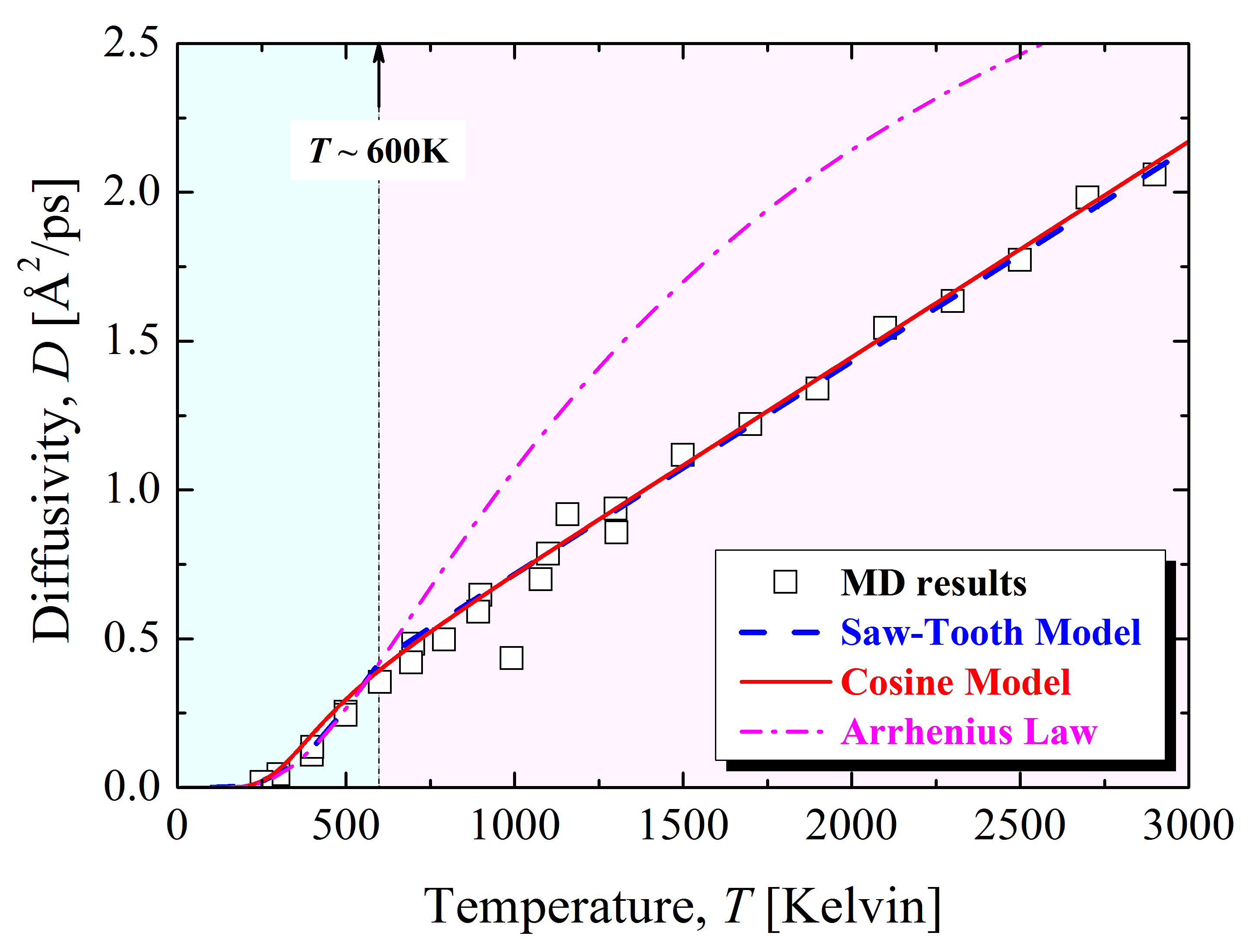}
			\caption{(Color online) The numerical results of temperature-dependent diffusivity (red solid-line) of single helium in BCC W estimated using Eq.~(\ref{Eq.:modified_D}), where the parameters are obtained from MD simulation in Ref.~\cite{wen_many-body_2017}, i.e., $D_0 = 4.3$\AA/ps, $E_m = 0.12 $eV, and $\mu = 8.4 \textrm{\AA}^2/\textrm{ps}/\textrm{eV}$, getting $b = 1.36$, as well as the comparison with MD simulation results (black open-square) \cite{wen_many-body_2017, perez2014diffusion}, the predictions of saw-tooth model with the same input parameters (blue dashed-line) \cite{wen_many-body_2017} and Arrhenius law (magnet dashed-dot-line) in Eq.~(\ref{Eq.:D_AR}). Left blue-shading and right red-shading regions denote the temperature ranges where helium diffusion reveals Arrhenius behavior and Einstein behavior, respectively, with the critical temperature being around 600K.}
			\label{Fig.3}
		\end{figure}
	Note that, parameters in Eq.~(\ref{Eq.:modified_D}) are all of many-body nature, which can be obtained from either atomistic modeling or experimental measurements. Using the parameters estimated in Ref.~\cite{wen_many-body_2017} for single helium diffusion in BCC W, i.e., $D_0 = 4.3$ \AA/ps, $E_m = 0.12 $eV, and $\mu = 8.4 \textrm{\AA}^2/\textrm{ps}/\textrm{eV}$, getting $b = 1.36$, the temperature dependence of diffusivity in Eq.~(\ref{Eq.:modified_D}) could be achieved. Here, this input migration energy $E_m = 0.12$eV, obtained by Arrhenius-fitting the dynamic simulation results at low-temperature limit, is a slightly smaller than that obtained in molecular static calculation (i.e., $E_m = 0.145$eV from MCG method), which might be arising from the entropic term, or the temperature dependence of migration energy. The results are plotted in Fig.~\ref{Fig.3}, as well as the comparison with MD simulation results \cite{wen_many-body_2017, perez2014diffusion} and the prediction using saw-tooth model \cite{wen_many-body_2017} and the Arrhenius equation (Eq.~(\ref{Eq.:D_AR})) with the same input parameters. It can be seen that our current model, i.e., \emph{cosine model} in Fig.~(\ref{Fig.3}),  could reproduce the feature of low-energy reaction at various temperatures ranging from $\sim$0K to 3000K. A slight difference in results between cosine model Eq.~(\ref{Eq.:modified_D}) and saw-tooth model appears in the low-temperature region, i.e., 300K $< T < 500$K, where the Arrhenius behavior is revealed, because of the usage of different potential to describe the characteristics of the equilibrium state for helium migration. On the other hand, the Arrhenius equation give rise to a relative large overestimation for diffusivity at high-temperature region, e.g., $T > 600$K. 
	
\begin{figure}
			\graphicspath{}
			\makeatletter
			\def\@captype{figure}
			\makeatother
			\centering
			\includegraphics[width=0.45\textwidth]{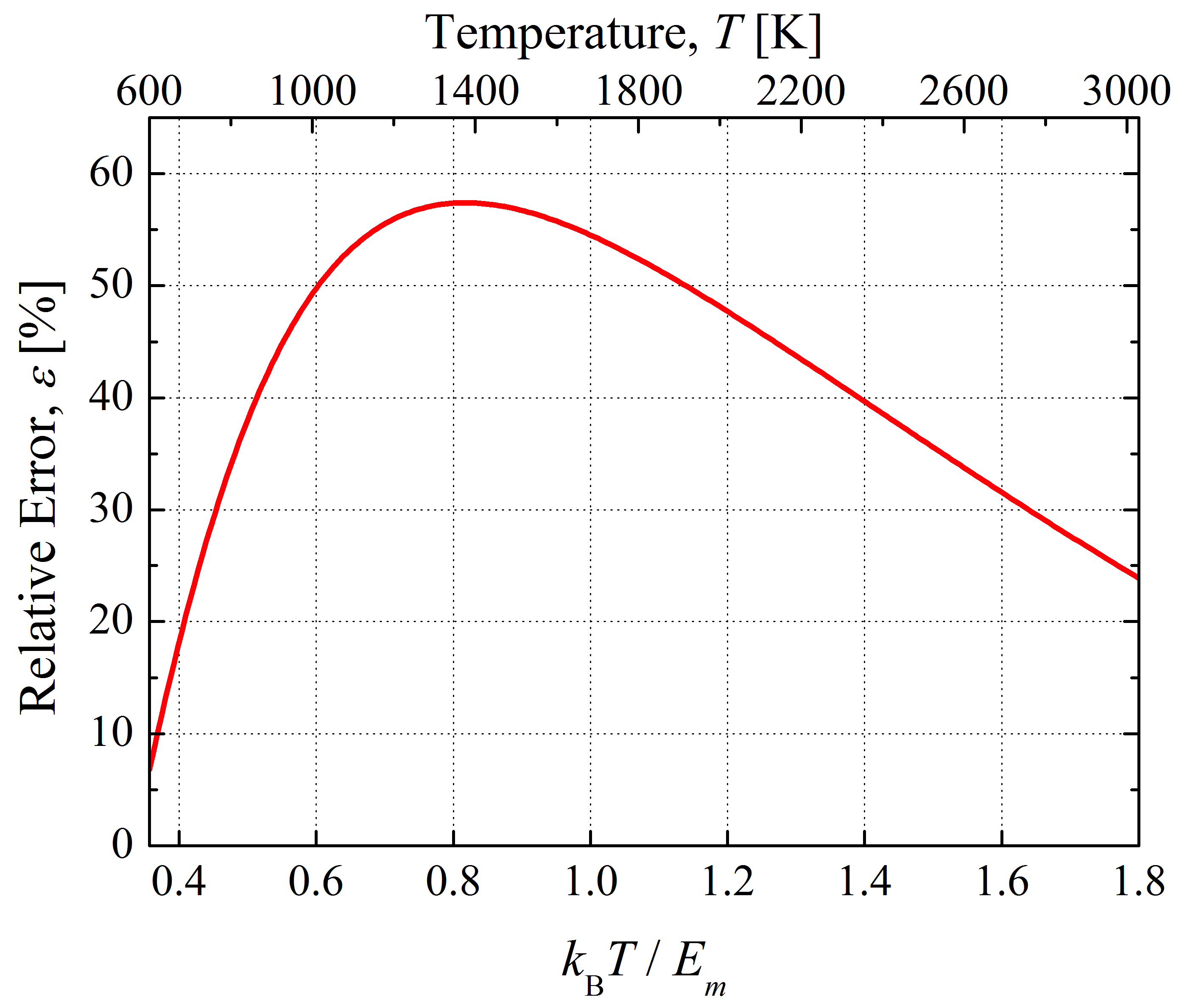}
			\caption{The defined relative error $\varepsilon$ of the prediction of diffusivity using Arrhenius equation (i.e., Eq.~(\ref{Eq.:D_AR})), compared to that using Einstein equation (i.e., Eq.~(\ref{Eq.:ED_equation})), in the high-temperature region, e.g., $T > 600$K.}
			\label{Fig.4}
		\end{figure}

	In a word, in the case of $E_m \ll k_\textrm{B}T$, Einstein equation is more appropriate to describe the diffusion behavior of helium in metals, due to its low migration energy barrier. A relative error $\varepsilon(T)$ could be defined as 
		\begin{equation}
			\varepsilon(T) = \frac{\left| D_A(T) - D_E(T)\right| }{D_E(T)} \times 100\%
		\end{equation} 
	to check the error of the application of Arrhenius equation in the high-temperature region, e.g., $T > 600$K in BCC W. Here, $D_A(T)$ and $D_E(T)$ are the diffusivity predicted using Arrhenius equation (i.e., Eq.~(\ref{Eq.:D_AR})) and Einstein equation (i.e., Eq.~(\ref{Eq.:ED_equation})), respectively. Fig.~(\ref{Fig.4}) presents the temperature dependence of $\varepsilon$. It can be seen $\varepsilon$ could be as large as 50\% at 1000K$< T < 2000$K, resulting in a significant error of growth rate of helium-vacancy clusters or bubbles, thus the kinetics of microstructural evolution in nuclear materials.
	
	\section{\label{Sec.4}Conclusion}
	For interstitial helium solute in metals, its non-Arrhenius diffusion behavior at finite temperatures could be completely described by the quasi-equilibrium diffusion parameters, e.g., pre-factor $D_0$ and migration energy $E_m$, in the framework of Arrhenius equation. The non-equilibrium diffusion parameter, e.g., classical mobility $\mu$, should be involved to describe the diffusion behavior at high-temperatures. In this paper, we proposed a stochastic model and derived a modified formula, with those three parameters involved, to describe the non-Arrhenius diffusion behavior of single interstitial helium in metals, which could help to develop a more accurate lattice based kinetic Monte Carlo scheme for the long-term evolution of helium-vacancy clusters and bubble. Our model is built up on the basis of a stochastic model of Brownian motion upon a cosine potential. We assumed that the helium migration could be treated as a step-by-step random hopping event so that the diffusivity satisfy the Lifson-Jackson's formula \cite{Lifson-Jackson} under overdamped condition. Further, by introducing an \emph{effective quality factor} to account for the damping feature, a modified formula for diffusivity (i.e., Eq.~(\ref{Eq.:modified_D})) or jump-frequency (i.e., Eq.~(\ref{Eq.:modified_nu})) of single helium solute is achieved. Of course, the story is not ending, because the related diffusion parameters still have to be obtained empirically or from other calculations. Compared to the saw-tooth model \cite{wen_many-body_2017}, the cosine potential used here is more comfortable to estimate the migratory information near the basin of potential well self-consistently, such as the attempt frequency, which could be presented in our further paper. It has to be noted that, our modified formula is not restricted in the case of helium diffusion, but could be promoted to the general cases of low-energy reaction in materials science. 
	
\section*{Acknowledgment}	
	This work is initiated and funded by the Guangzhou Science and Technology Project (No. 201707020002) and NSFC (No. 11672339, No. 11602310, No 11602311), to which the authors are thankful. Y. Zheng also thanks support from the Special Program for Applied Research on Super Computation of the NSFC-Guangdong Joint Fund (the second phase), Fok Ying Tung Foundation, Guangdong Natural Science Funds for Distinguished Young Scholar and China Scholarship Council.	
	
\section*{Reference}
\bibliographystyle{elsarticle-num}
\bibliography{He_Mig}

\end{document}